\documentclass{article}
\usepackage{amssymb}
\newtheorem{theorem}{Theorem}

\newtheorem{corollary}[theorem]{Corollary}

\newtheorem{definition}[theorem]{Definition}

\newtheorem{proposition}[theorem]{Proposition}

\newenvironment{proof}[1][Proof]{\noindent\textbf{#1.} }{\ \rule{0.5em}{0.5em}}
\begin{document}
\title{The essential spectrum of Schr\"{o}dinger operators on lattices}
\author{V. S. Rabinovich\thanks{The authors are grateful for
the support by the German Research Foundation (DFG), grant
444 MEX-112/2/05.}, S. Roch}
\date{}
\maketitle
\begin{abstract}
The paper is devoted to the study of the essential spectrum of 
discrete Schr\"{o}dinger operators on the lattice $\mathbb{Z}^{N}$ 
by means of the limit operators method. This method has been applied 
by one of the authors to describe the essential spectrum of 
(continuous) electromagnetic Schr\"{o}dinger operators, square-root 
Klein-Gordon operators, and Dirac operators under quite weak assumptions 
on the behavior of the magnetic and electric potential at infinity. 
The present paper is aimed to illustrate the applicability and 
efficiency of the limit operators method to discrete problems as 
well.

We consider the following classes of the discrete Schr\"{o}dinger
operators: 1) operators with slowly oscillating at infinity
potentials, 2) operators with periodic and semi-periodic potentials; 
3) Schr\"{o}dinger operators which are discrete quantum analogs of 
the acoustic propagators for waveguides; 4) operators with potentials 
having an infinite set of discontinuities; and 5) three-particle 
Schr\"{o}dinger operators which describe the motion of two particles 
around a heavy nuclei on the lattice $\mathbb{Z}^3$.
\end{abstract}
\section{Introduction}
The present paper deals with applications of the limit operators
method to the description of the essential spectrum of several
classes of discrete Schr\"{o}dinger operators. In \cite{RJMP}, 
this method has been applied to study the location of the essential
spectrum of electromagnetic Schr\"{o}dinger operators, square-root
Klein-Gordon operators, and Dirac operators under very weak assumptions 
on the behavior of magnetic and electric potentials at infinity. 
One remarkable outcome of this approach is a simple and transparent 
proof of the well known Hunziker, van Winter, Zjislin (HWZ) theorem 
on the location of the essential spectrum of multi-particle Hamiltonians 
(\cite{Cycon}).

Roughly speaking, the limit operators approach of \cite{RJMP} 
works as follows. The study of the essential spectrum of an unbounded 
operator is reduced to the study of the essential spectrum of a 
related bounded operator which belongs to a certain Banach 
algebra $\mathcal{B}$. To each operator $A \in \mathcal{B}$, there
is associated a family $\{ A_h \}$ of operators, called the limit 
operators of $A$, which reflect the behaviour of the operator $A$ 
at infinity. Based on results of \cite{RRS1,RRS2}, it is shown in 
\cite{RJMP} that
\begin{equation} \label{0}
sp_{ess} \, (A) = \bigcup sp \, (A_h)  
\end{equation}
where the union is taken over all limit operators $A_h$ of $A$. 
In general, the limit operators of a given operator have a simpler 
structure than the operator itself. Hence, in many instances,  
(\ref{0}) provides an effective tool for the description of the 
essential spectrum of operators in $\mathcal{B}$. An identity 
similar to (\ref{0}) holds for operators in the Wiener algebra on 
$\mathbb{Z}^{N}$ (see below).

A formula similar to (\ref{0}) has been obtained independently 
(but later) in the recent paper \cite{LastSimon} by using a 
localization technique and an appropriate partition of unit on a 
Hilbert space, and in \cite{AMP,Gorgescu1,Gorgescu2,Mantoiu} (see 
also the references cited therein) by means of $C^*$-algebra 
techniques. Note that the methods used in  
\cite{AMP,Gorgescu1,Gorgescu2,LastSimon,Mantoiu} are restricted to 
the description of the essential spectrum of {\em selfadjoint or 
normal} operators acting on a Hilbert space whereas the limit 
operators approach which will be discussed below works for non-selfadjoint 
operators acting on $L^{p}$-type spaces, for example for Schr\"{o}dinger 
operators with complex potentials on $L^p$-spaces, as well.

In this paper, we consider discrete electromagnetic Schr\"{o}\-dinger 
operators of the form
\begin{equation} \label{01}
Hu = \sum_{k=1}^N \frac{1}{2m_k} (V_{e_k}-a_k I) (V_{-e_k}
- \bar{a}_k I) u + \Phi u 
\end{equation}
on the lattice $\mathbb{Z}^N$. Here, $(V_g u)(x) := u(x-g), \, x \in 
\mathbb{Z}^N$, is the operator of shift by $g \in \mathbb{Z}^N$, the 
$e_j$ are the vectors $(0, \, \ldots, \, 1, \, \ldots, \, 0)$ with 
the $1$ standing at the $j$th position, $m_k$ is the mass of the 
$k$th particle, and the $a_k$, $k = 1, \, \ldots, \, N$ and $\Phi$ are 
bounded complex-valued functions on $\mathbb{Z}^N$. The vector-valued 
function $a=(a_1, \, \ldots, \, a_N)$ can be considered as an analogue 
of the magnetic potential, whereas $\Phi$ is the discrete analogue of 
the electric potential. If $\Phi$ is real-valued, then $H$ is a 
selfadjoint operator on the Hilbert space $l^2 (\mathbb{Z}^N)$. 

Operators of the form (\ref{01}) describe the so-called tight
binding model in solid state physics (see \cite{Mattis,Mogilner}
and the references given there), which plays a prominent role in the
theory of propagation of spin waves and of waves in quasi-crystals
\cite{Teschl,SR3}, in the theory of nonlinear integrable lattices
\cite{Teschl,Deift}, and in other places. There is an extensive 
bibliography devoted to different aspects of the spectral theory 
of discrete Schr\"{o}dinger operators. Let us only mention
\cite{AlbaverioLakaev,LakaevMuminov,AlbLakMum,Shubindis,LastSimon,
Teschl,Yafaev}, and see also the references cited in these papers.

The present paper is organized as follows. In Section 2, we recall
some auxiliary material concerning the Wiener algebra of discrete
operators on $\mathbb{Z}^N$ and the limit operators method. Our
presentation will follow the papers \cite{RRS1,RRS2}. A
comprehensive account of this material can be found in 
\cite{RRSB}. In Section 3, we give applications of the limit
operators method to the description of the essential spectrum of 
the discrete Schr\"{o}dinger operator (\ref{01}) under different 
assumptions on the behavior of magnetic and electric potentials
at infinity. In particular, we are going to describe the essential 
spectra for the following classes of potentials.

1) {\em Slowly oscillating at infinity potentials.} This class of
potentials is remarkable, since the limit operators of 
Schr\"{o}dinger operators with slowly oscillating potentials are 
unitarily equivalent to operators of the form
\begin{equation} \label{02}
\sum_{k=1}^N \frac{1}{2m_k} \Delta_k + \Phi_h I,
\end{equation}
where $\Phi_h$ is constant and $\Delta_k$ is the discrete
Laplacian $2I - V_{e_k}-V_{-e_k}$  with respect to the variable 
$x_k \in \mathbb{Z}$.

2) {\em Periodic and semi-periodic potentials.} For periodic
operators, the essential spectrum coincides with the spectrum.
There exists a matrix realization of (\ref{01}) which allows one 
to describe the spectrum explicitly. We prove that the limit
operators for Schr\"{o}dinger operators with semi-periodic
potentials are periodic operators. Another approach to
one-dimensional periodic Jacobi operators is given in
\cite{Teschl}.

3) {\em Quantum waveguides.} We consider the discrete
Schr\"{o}dinger operator $H = \sum_{k=1}^N \Delta_k + \Phi I$
which is a discrete quantum analog of the acoustic propagator for
waveguides (see \cite{Weder,Wilcox}). The limiting absorption
principle as well as the scattering theory for the discrete
acoustic propagator are considered in \cite{RabinD}.

4) {\em Potentials with an infinite set of discontinuities.} We
study the one-dimen\-sion\-al Schr\"{o}dinger operator with
electric potential $\Phi$ which takes two real values $a$ and $b$
and which has a countable set of discontinuities. More precisely,
$\Phi$ is equal to $a$ on  $\Lambda = \cup _{k=0}^\infty \{ x \in
\mathbb{Z} : \gamma_k^- \le |x| \le \gamma_k^+ \}$ and equal to
$b$ on $\mathbb{R} \setminus \Lambda$ where
\begin{equation} \label{04'}
\lim_{k \to \infty} \gamma_k^- = \lim_{k \to \infty} \gamma_k^+ =
\lim_{k \to \infty} (\gamma_k^+ - \gamma_k^-) = \lim_{k \to
\infty} (\gamma_k^- - \gamma_{k-1}^+) = +\infty.
\end{equation}
We prove that $sp_{ess} \, (\Delta + \Phi I) = [a, \, a+4] \cup [b,
\, b+4]$ in this case.

5) We consider the three-particle Schr\"{o}dinger operator $H$ on
$l^2 (\mathbb{Z}^6)$ which describes the motion of two particles
$x^1, \,  x^2 \in \mathbb{Z}^3$ with masses $m_1, \, m_2$ on the
lattice $\mathbb{Z}^3$ around a heavy nuclei located at the point
$0$. We prove that
\begin{equation} \label{04}
sp_{ess} \, (H) = sp \, (H_1) \cup sp \, (H_2) \cup sp \, (H_{12}),
\end{equation}
where $H_j, \, j=1, \, 2$ is the Hamiltonian of the subsystem 
consisting of the particle $x^j$ and the nuclei, and where $H_{12}$ 
is the Hamiltonian of the subsystem consisting of $x^1$ and $x^2$. 
We further apply (\ref{04}) to estimate the lower and upper bound 
of the essential spectrum of the three-particle Hamiltonian. Compare 
also the papers \cite{AlbaverioLakaev,AlbLakMum,LakaevMuminov} 
devoted to spectral properties of three-particle problems in the 
impulse representation.
\section{Auxiliary material}
\subsection{Fredholm theory and essential spectrum of operators in
the Wiener algebra}
We will use the following notations. For each Banach space $X$,
let $L(X)$ denote the Banach algebra of all bounded linear
operators acting on $X$ and $K(X)$ the ideal of $L(X)$ of all
compact operators. An operator $A \in L(X)$ is called a {\em
Fredholm operator} if $\mbox{ker} \, A := \{x \in X : Ax = 0\}$
and $\mbox{coker} \, A := X/A(X)$ are finite dimensional spaces.
The essential spectrum of $A$ consists of all complex numbers
$\lambda$ for which the operator $A - \lambda I : X \to X$ is not
Fredholm. We denote the spectrum and the essential spectrum of $A
\in L(X)$ by $sp \, (A | X)$ and $sp_{ess} \, (A | X)$,
respectively. If the dependence of these spectra on $X$ is evident, 
we simply write $sp \, (A)$ and $sp_{ess} \, (A)$ for the spectrum 
and the essential spectum of $A$. The discrete spectrum of $A$,
denoted by $sp_{dis} \, (A | X)$ or $sp_{dis} \, (A)$, is the set 
of all isolated eigenvalues of finite multiplicity of $A$. Note 
that
\[
sp_{dis} \, (A | X) \subseteq sp \, (A | X) \setminus 
sp_{ess} \, (A | X)
\]
and that equality holds in this inclusion if $A$ is self-adjoint. 
Moreover, if $D \subseteq \mathbb{C} \setminus sp_{ess} \, (A)$
is a domain which contains at least one point which is not in 
$sp (A)$, then
\[
sp \, (A) \cap D = sp_{dis} \, (A) \cap D
\]
(Corollary 8.4 in Chapter XI of \cite{GGK1}).  

For $p \in [1, \, \infty)$ and $N$ a positive integer, let 
$l^p(\mathbb{Z}^N)$ denote the Banach space of all functions $u : 
\mathbb{Z}^N \to \mathbb{C}$ for which the
norm
\[
\|u\|_{l^p(\mathbb{Z}^N)} := \left( \sum_{x \in \mathbb{Z}^N} |u(x)|^p
\right)^{1/p} 
\]
is finite, and let $l^\infty (\mathbb{Z}^N)$ refer to the $C^*$-algebra
of all bounded complex-valued functions on $\mathbb{Z}^N$ with norm
\[
\|a\|_{l^\infty (\mathbb{Z}^N)} := \sup_{x \in \mathbb{Z}^N} |a(x)|.
\]
The following results hold for operators on $l^p$-spaces of 
vector-valued sequences and (with modifications) for operators 
on $l^p$-spaces of sequences with values in an arbitrary Banach space
as well. We will not need these results here and refer the 
monograph \cite{RRSB} for details.

Let $\{a_\alpha \}_{\alpha \in \mathbb{Z}^N}$ be a sequence of 
functions in $l^\infty (\mathbb{Z}^N)$. Consider the difference 
operator
\begin{equation} \label{2.1}
(Au)(x) := \sum_{\alpha \in \mathbb{Z}^N} a_\alpha (x) (V_\alpha
u) (x), \quad x \in \mathbb{Z}^N,
\end{equation}
which is well defined for compactly supported functions $u$. We let 
$W(\mathbb{Z}^N)$ stand for the class of all operators of the form 
(\ref{2.1}) which satisfy
\begin{equation} \label{2.2}
\|A\|_{W(\mathbb{Z}^N)} := \sum_{\alpha \in \mathbb{Z}^N} 
\|a_\alpha\|_{l^\infty(\mathbb{Z}^N)} < \infty.
\end{equation}
The set $W(\mathbb{Z}^N)$ provided with standard operations and 
with the norm (\ref{2.2}) forms a Banach algebra, the so-called 
{\em Wiener algebra} on $\mathbb{Z}^N$. It is easy to see that 
this algebra is continuously embedded into $L(l^p (\mathbb{Z}^N))$ 
for each $p \in [1, \, \infty]$.
\begin{proposition} \label{p2.1}
Let $p \in [1, \, \infty]$. The Wiener algebra $W(\mathbb{Z}^N)$ 
is inverse closed in each of the algebras $L(l^p(\mathbb{Z}^N))$, 
that is, if $A \in W(\mathbb{Z}^N)$ possesses a bounded inverse 
$A^{-1}$ on $l^p (\mathbb{Z}^N)$, then $A^{-1} \in W(\mathbb{Z}^N)$.
\end{proposition}
\begin{proposition} \label{c2.1}
Let $A \in W(\mathbb{Z}^N)$. Then the spectrum $sp \, (A | 
l^p(\mathbb{Z}^N))$ does not depend on $p \in [1, \, \infty]$.
\end{proposition}
The following definition introduces our main tool to describe the 
essential spectra of operators in the Wiener algebra.    
\begin{definition} \label{d2.1}
Let $p \in (1, \, \infty)$, and let $h : \mathbb{N} \to
\mathbb{Z}^N$ be a sequence tending to infinity. An operator
$A_h \in L(l^p (\mathbb{Z}^N))$ is called the {\em limit operator 
of} $A \in L(l^p (\mathbb{Z}^N))$ {\em defined by the sequence} $h$ 
if the strong limits on $l^p (\mathbb{Z}^N)$
\[
A_h = \mbox{\rm s-lim}_{j \to \infty} V_{-h(j)} A V_{h(j)}, \quad
A_{h}^* = \mbox{\rm s-lim}_{j \to \infty} V_{-h(j)} A^* V_{h(j)}
\]
exist. We write $op \, (A | l^p(\mathbb{Z}^N))$ or simply 
$op \, (A)$ for the set of all limit operators of $A$, and we call 
this set the {\em operator spectrum of} $A$.
\end{definition}
Let the sequence $h : \mathbb{N} \to \mathbb{Z}^N$ tend to
infinity, and let $A \in W (\mathbb{Z}^N)$ be as in (\ref{2.1}). Then
\[
(V_{-h(j)} A V_{h(j)} u)(x) = \sum_{\alpha \in \mathbb{Z}^N}
a_\alpha (x + h(j)) (V_\alpha u)(x).
\]
Applying the Weierstrass-Bolzano theorem and a Cantor diagonal
process, one gets a subsequence $g$ of $h$ such that, for every 
$\alpha \in \mathbb{Z}^N$, there is a function $a_\alpha^g$ with
\[
a_\alpha (x + g(j)) \to a_\alpha^g(x), \quad x \in \mathbb{Z}^N.
\]
Moreover,
\begin{equation} \label{2.3}
\|a_\alpha^g\|_{l^\infty (\mathbb{Z}^N)} \le \|a_\alpha\|_{l^\infty 
(\mathbb{Z}^N)}.
\end{equation}
Set
\[
A_g := \sum_{\alpha \in \mathbb{Z}^N} a_\alpha^g V_\alpha.
\]
It follows from (\ref{2.3}) that $A_g$ belongs to the Wiener algebra 
$W(\mathbb{Z}^N)$ again, and it has been proved in \cite{RRS1} that 
$A_g$ is indeed the limit operator of $A$ defined by the sequence $g$.
\begin{proposition} \label{c2.1a}         
The operator spectrum $op \, (A | l^p(\mathbb{Z}^N))$ does not depend 
on $p \in (1, \, \infty)$ for $A \in W(\mathbb{Z}^N)$.
\end{proposition}         
\begin{theorem} \label{t2.0}
Let $A \in W(\mathbb{Z}^N)$ and $p \in (1, \, \infty)$. Then $A : 
l^p(\mathbb{Z}^N) \to l^p(\mathbb{Z}^N)$ is a Fredholm operator if 
and only if there exists a $p_0 \in [1, \, \infty]$ such that all
limit operators of $A$ are invertible on $l^{p_0} (\mathbb{Z}^N)$.
\end{theorem}
In particular this shows that the invertibility of all limit
operators of $A$ on $l^{p_0} (\mathbb{Z}^N)$ for
some $p_0$ implies their invertibility on $l^p (\mathbb{Z}^N)$ 
for every $p$.

The following theorem is an important corollary of Theorem
\ref{t2.0}.
\begin{theorem} \label{t2.1}
Let $A \in W(\mathbb{Z}^N)$. Then, for every $p \in (1, \, \infty)$,
\begin{equation} \label{2.4}
sp_{ess} \, (A | l^p(\mathbb{Z}^N)) = 
\bigcup_{A_g \in op \, (A |l^p(\mathbb{Z}^N))} 
sp \, (A_g | l^p(\mathbb{Z}^N)).
\end{equation}
\end{theorem}
According to Propositions \ref{c2.1} and \ref{c2.1a}, neither the 
spectra $sp \, (A_g | l^p(\mathbb{Z}^N))$ nor the operator 
spectrum $op \, (A | l^p(\mathbb{Z}^N))$ do depend on $p \in (1, \, 
\infty)$. Hence, the following holds.
\begin{theorem} \label{t2.2}
Let $A \in W(\mathbb{Z}^N)$. Then the essential spectrum of the 
operator $A$, considered as acting on $l^p(\mathbb{Z}^N)$, does not 
depend on $p \in (1, \, \infty)$.
\end{theorem}
This theorem and the above remarks justify to write the equality 
(\ref{2.4}) simply as 
\begin{equation} \label{2.4'}
sp_{ess} \, (A) = \bigcup_{A_g \in op \, (A)} sp \, (A_g).
\end{equation}
The following proposition gives a sufficient condition for the
absence of the discrete spectrum of an operator $A$ in the Wiener 
algebra.
\begin{proposition} \label{pr2.1}
Let $A \in W(\mathbb{Z}^N)$, and assume there exist a sequence 
$h \to \infty$ and an operator $A_h$ such that
\begin{equation} \label{e2.5}
\lim_{j \to \infty} \|V_{-h(j)} A V_{h(j)} - A_h\| = 0.
\end{equation}
Then $sp_{ess} \, (A) = sp \, (A)$.
\end{proposition}
\begin{proof}
Let $\lambda \notin sp_{ess} \, (A)$. Then $A - \lambda I$ is a
Fredholm operator. By Theorem \ref{t2.0}, all limit operators $A_h
- \lambda I$ are invertible. Hence, $\lambda \notin sp \, (A_h)$ for
every limit operator $A_h$. It follows from (\ref{e2.5}) that
$\lambda \notin sp \, (A)$. Hence, $sp \, (A) \subseteq sp_{ess} \,
(A)$.
\end{proof}
\subsection{Slowly oscillating coefficients}
A function $a \in l^\infty (\mathbb{Z}^N)$ is said to be {\em 
slowly oscillating at infinity} if
\[
\lim_{x \to \infty} (a(x+y) - a(x)) = 0
\]
for every $y \in \mathbb{Z}^N$. We denote the class of all slowly
oscillating matrix-functions by $SO (\mathbb{Z}^N)$. If $a$ is 
slowly oscillating then, for every sequence $h \to \infty$, there 
are a subsequence $g$ of $h$ and a complex number $a^g$ such that
\[
\lim_{j \to \infty} a(x+g(j)) = \lim_{j \to \infty} a(g(j)) = a^g
\]
for each $x \in \mathbb{Z}^N$ (see \cite{RRS1}). Thus, every limit 
function of a slowly oscillating function is constant. We write 
$W^{SO} (\mathbb{Z}^N)$ for the subalgebra of $W(\mathbb{Z}^N)$ 
which contains all operators with slowly oscillating coefficients. 
If $A \in W^{SO} (\mathbb{Z}^N)$, then all limit operators of $A$ 
have the form
\begin{equation} \label{2.5}
A^g := \sum_{\alpha \in \mathbb{Z}^N} a_\alpha^g V_\alpha, \quad
\mbox{with} \; a_\alpha^g \in \mathbb{C},
\end{equation}
i.e., they are operators with constant diagonals.

For a complex-valued function $u$ on $\mathbb{Z}^N$ with compact 
support, we define their discrete Fourier transform  $Fu = \hat{u}$ 
by
\[
\hat{u}(t) = (Fu)(t) := \frac{1}{(2 \pi)^{N/2}} \sum_{x \in
\mathbb{Z}^N} u(x)t^{-x}, \quad t \in \mathbb{T}^N,
\]
where $\mathbb{T}$ is the unit circle on the complex plane, 
considered as a multiplicative group, $\mathbb{T}^N := 
\mathbb{T} \times \ldots \times \mathbb{T}$ is the $N$-dimensional 
torus, and $t^x := t_1^{x_1} \ldots t_N^{x_N}$ for $t \in \mathbb{T}^N$ 
and $x \in \mathbb{Z}^N$. Further, let
\[
d\mu (t) := \frac{1}{(2\pi)^{N/2} \, i^N} \frac{dt_1 \ldots dt_N}{t_1
\ldots t_N}
\]
be the normalized Haar measure on $\mathbb{T}^N$. Then
the inverse Fourier transform is given by
\[
u(x) = (F^{-1} \hat{u})(x) = \int_{\mathbb{T}^N} \hat{u}(t) \, t^x 
\, d\mu (t), \quad x \in \mathbb{Z}^N.
\]
Moreover, the Fourier transform extends continuously to a unitary
operator $F : l^2(\mathbb{Z}^N) \to L^2 (\mathbb{T}^N, \, d\mu)$ 
which has $F^{-1}$ as its adjoint.

The discrete Fourier transform establishes a unitary equivalence 
between the operator $A^g$ in (\ref{2.5}) and the operator of
multiplication by the function
\[
\hat{A}^g : \mathbb{T}^N \to \mathbb{C}, \quad t \mapsto 
\sum_{\alpha \in \mathbb{Z}^N} a_\alpha^g \, t^{-\alpha}.
\]
In combination with (\ref{2.4'}), this yields the following.
\begin{theorem}
Let $A \in W^{SO} (\mathbb{Z}^N)$. Then
\[
sp_{ess} \, (A) = \bigcup_{A_g \in op \, (A)} sp \, (A_g) = 
\bigcup_{A_g \in op \, (A)} \bigcup_{t \in \mathbb{T}^N} 
\{ \hat{A}^g (t) \}.
\]
\end{theorem}
\subsection{Almost-periodic, $r$-periodic, and semi-periodic 
coeffi\-cients}
A function $a \in l^\infty (\mathbb{Z}^N)$ is said to be
{\em almost-periodic} if, for each sequence $h$ tending to infinity,
there are a subsequence $g$ of $h$ and a function $a^g \in
l^\infty (\mathbb{Z}^N)$ such that
\[
\lim_{j \to \infty} \|V_{-g(j)} a - a^g\|_{l^\infty
(\mathbb{Z}^N)} = 0.
\]
The almost-periodic functions form a $C^*$-subalgebra of $l^\infty 
(\mathbb{Z}^N)$ which we denote by $AP(\mathbb{Z}^N)$. We further 
write $W^{AP} (\mathbb{Z}^N)$ for the subalgebra of $W(\mathbb{Z}^N)$ 
which contains all operators with almost-periodic coefficients.
\begin{proposition} \label{prop2.1}
Let $A \in W^{AP} (\mathbb{Z}^N)$, and let $A_h$ be the limit
operator of $A$ defined by a sequence $h \to \infty$. Then there
is a subsequence $g$ of $h$ such that
\begin{equation} \label{e2.4}
\lim_{j \to \infty} \|V_{-g(j)} A V_{g(j)} - A_h\| =0.
\end{equation}
\end{proposition}
Thus, $A_h$ is a limit operator with respect to {\em norm convergence}.
The proof follows immediately from the definition of the algebra
$W^{AP} (\mathbb{Z}^N)$.
\begin{corollary} \label{cor2.1}
Let $A \in W^{AP} (\mathbb{Z}^N)$, and let $A_h$ be a limit operator
of $A$. Then
\begin{equation} \label{f2.4}
sp \, (A) = sp_{ess} \, (A) \quad \mbox{and} \quad sp \, (A) = 
sp \, (A_h).
\end{equation}
\end{corollary}
Indeed, the first equality follows from Proposition \ref{pr2.1}, 
and the second one is a consequence of (\ref{e2.4}). \\[3mm]
Let $r$ be an $N$-tuple of positive integers. A function $a$ on 
$\mathbb{Z}^N$ is called {\em $r$-periodic} if $a(x + r) = a(x)$ 
for each $x \in \mathbb{Z}^N$. The class of all $r$-periodic 
functions is denoted by $P_r (\mathbb{Z}^N)$. We further write 
$W^{P_r} (\mathbb{Z}^N)$ for the subalgebra of $W(\mathbb{Z}^N)$ 
which consists of all operators with $r$-periodic coefficients. 
Evidently, 
\[
W^{P_r} (\mathbb{Z}^N) \subset W^{AP} (\mathbb{Z}^N).
\]
Hence, the equalities (\ref{f2.4}) hold for operators in $W^{P_r} 
(\mathbb{Z}^N)$ as well.

Let $r := (r_1, \, \ldots, \, r_N)$ and $d := r_1 \cdot \ldots 
\cdot r_N$, and write $\mathbb{C}^d$ as
\[
\mathbb{C}^{r_1} \otimes \ldots \otimes \mathbb{C}^{r_N} =:
\otimes_{j=1}^N \mathbb{C}^{r_j}.
\]
Let $\{ e_1, \, \ldots, \, e_N \}$ stand for the standard basis of
$\mathbb{C}^N$, i.e., the $j$th entry of $e_j$ is one, and the
other entries are zero. Then the vectors $\{e_{j_1} \otimes \ldots
\otimes e_{j_N} \}_{j_1=1, \, \ldots , \, j_N=1}^{r_1, \, \ldots,
\, r_N}$ form a basis of $\mathbb{C}^d = \otimes_{j=1}^N
\mathbb{C}^{r_j}$. We put these vectors into lexicographic order. 
Let
\[
T_r : l^2(\mathbb{Z}^N) \to l^2(\mathbb{Z}^N, \, (\otimes_{j=1}^N
\mathbb{C}^{r_j}))
\]
be the unitary operator defined by
\begin{equation} \label{f2.5}
(T_r f) (y_1, \, \ldots, \, y_N) := (f(r_1 y_1 +j_1-1, \, \ldots, \,
r_N y_N + j_N-1))_{j_1=1, \, \ldots, \, j_N=1}^{r_1, \, \ldots, \, r_N}.
\end{equation}
For $a \in l^\infty (\mathbb{Z}^N)$, set $\mu (a) := T_r a T_r^{-1}$.
Clearly, $\mu(a)$ is given by the diagonal matrix
\begin{equation} \label{2.5'}
(\mbox{diag} \, (a (j_1, \, \ldots, \, j_N))_{j_1=1, \, \ldots,
\, j_N=1}^{r_1, \, \ldots, \, r_N}
\end{equation}
with the entries on the main diagonal being in lexicographic order.
Further, one has
\[
T_r (V_1^{\alpha_1} \ldots V_N^{\alpha_N}) T_r^{-1} =
\Lambda_1^{-\alpha_1} \otimes \ldots \otimes \Lambda_N^{-\alpha_N}
\]
where $\Lambda_j$ is the $r_j \times r_j$-matrix operator
\begin{equation} \label{f2.6}
\Lambda _{j}=\left(
\begin{array}{ccccc}
0 & I & 0 & \cdot & 0 \\
0 & 0 & I & \cdot & 0 \\
\cdot & \cdot & \cdot & \cdot & \cdot \\
0 & 0 & 0 & \cdot & I \\
V_{j}^{-1} & 0 & 0 & \cdot & 0
\end{array}
\right).
\end{equation}
Hence, the operator $A := \sum_{\alpha \in \mathbb{Z}^N} a_\alpha
V_\alpha \in W^{P_r} (\mathbb{Z}^N)$ acting on $l^2 (\mathbb{Z}^N)$
is unitarily equivalent to the operator
\[
\mathcal{A} := T_r A T_r^{-1} = \sum_{\alpha \in \mathbb{Z}^N}
\mu (a_\alpha) \, \Lambda_1^{-\alpha_1} \otimes \ldots \otimes
\Lambda_N^{-\alpha_N}
\]
acting on $l^2 (\mathbb{Z}^N, \, \otimes_{j=1}^N \mathbb{C}^{r_j})$.
Furthermore, the operator $\mathcal{A}$ is unitarily equivalent to
the operator of multiplication by the continuous function
$\sigma (\mathcal{A}) : \mathbb{T}^N \to \otimes_{j=1}^N
\mathbb{C}^{r_j}$ which maps $t = (t_1, \, \ldots, \, t_N) \in
\mathbb{T}^N$ to
\begin{equation} \label{2.6}
\sigma (\mathcal{A})(t) := \sum_{\alpha = (\alpha_1, \, \ldots, \,
\alpha_N) \in \mathbb{Z}^N} \mu (a_\alpha) \, \hat{\Lambda}_1^{-\alpha_1}
(t_1) \otimes \ldots \otimes \hat{\Lambda}_N^{-\alpha_N}(t_N)
\end{equation}
where
\[
\hat{\Lambda}_j (t_j) := \left(
\begin{array}{ccccc}
0 & 1 & 0 & \cdot & 0 \\
0 & 0 & 1 & \cdot & 0 \\
\cdot & \cdot & \cdot & \cdot & \cdot \\
0 & 0 & 0 & \cdot & 1 \\
t_j^{-1} & 0 & 0 & \cdot & 0
\end{array}
\right).
\]
\begin{theorem} \label{t2.3}
Let $A \in W^{P_r} (\mathbb{Z}^N)$ and $p \in [1, \, \infty]$.
The operator $A : l^p (\mathbb{Z}^N) \to l^p(\mathbb{Z}^N)$ is
invertible if and only if
\[
\det \sigma (\mathcal{A})(t) \neq 0 \quad \mbox{for each} \;
t \in \mathbb{T}^N.
\]
\end{theorem}
\begin{corollary} \label{co2.1}
Let $A \in W^{P_r} (\mathbb{Z}^N)$ and $p \in [1, \, \infty]$.
Then both the spectrum and the essential spectrum of $A :
l^p (\mathbb{Z}^N) \to l^p(\mathbb{Z}^N)$ are independent of $p$,
and
\[
sp_{ess} \, (A) = sp \, (A) = \cup_{t \in \mathbb{T}^N} sp \,
(\sigma (\mathcal{A} (t))).
\]
\end{corollary}
Let finally $SP_r (\mathbb{Z}^N)$ denote the smallest closed 
subalgebra of $l^\infty (\mathbb{Z}^N)$ which contains the algebras 
$SO (\mathbb{Z}^N)$ of the slowly oscillating functions and 
$P_r (\mathbb{Z}^N)$ of the $r$-periodic functions. Evidently, 
$SP_r (\mathbb{Z}^N)$ is a $C^*$-subalgebra of $l^\infty 
(\mathbb{Z}^N)$. We refer to the functions in this algebra as 
{\em semi-periodic functions}. Further, we write $W^{SP_r} 
(\mathbb{Z}^N)$ for the subalgebra of $W(\mathbb{Z}^N)$ of all 
operators with coefficients in $SP_r (\mathbb{Z}^N)$. 
\begin{proposition} \label{p2.3}
All limit operators of operators in $W^{SP_r} (\mathbb{Z}^N)$ 
belong to the algebra $W^{P_r} (\mathbb{Z}^N)$.
\end{proposition}
\begin{proof}
A Cantor diagonal argument shows that it is sufficient to prove
the assertion for diagonal operators in $W^{SP_r} (\mathbb{Z}^N)$,
i.e., for operators of multiplication by functions in $SP_r
(\mathbb{Z}^N)$. Another application of Cantor's diagonal argument
implies furthermore that it is sufficient to prove the assertion 
for a dense subset of $SP_r (\mathbb{Z}^N)$.

Thus, let $a = \sum_{k=1}^m a_k b_k$ with $a_k \in P_r (\mathbb{Z}^N)$
and $b_k \in SO (\mathbb{Z}^N)$, and let $h$ be a sequence tending 
to infinity. It is easy to see that there is a subsequence $l$ of $h$ 
such that
\[
V_{-l(j)} a_k V_{l(j)} = a_k \quad \mbox{for all} \; j \; \mbox{and} 
\; k.
\]
Further, by what has already been said, there are a subsequence 
$g$ of $l$ and constants $b_k^g$ such that
\[
\mbox{s-lim}_{j \to \infty} V_{-g(j)} b_k V_{g(j)} = b_k^g I \quad 
\mbox{for each} \; k.
\]
Thus, $\mbox{s-lim}_{j \to \infty} V_{-g(j)} a V_{g(j)}$ is equal to
$a^g := \sum_{k=1}^m a_k b_k^g$ which is in $P_r (\mathbb{Z}^N)$.
\end{proof}

\begin{proposition} \label{p2.4}
Let $A \in W^{SP_r} (\mathbb{Z}^N)$ and $p \in [1, \, \infty]$.
Then the essential spectrum of $A : l^p (\mathbb{Z}^N) \to
l^p(\mathbb{Z}^N)$ is independent of $p$, and
\[
sp_{ess} \, (A) = \bigcup_{A_h \in op \, (A)} \bigcup_{t \in 
\mathbb{T}^N} sp \, (\sigma (\mathcal{A}_h (t))).
\]
\end{proposition}
\section{Schr\"{o}dinger operators}
The aim of this section is to describe the essential spectrum of 
discrete Schr\"{o}\-dinger operators of the form
\begin{equation} \label{4.1}
H := \sum_{k=1}^N (V_{e_k} - a_k I) \, (V_{-e_k} - \bar{a}_k I) + \Phi I
\end{equation}
where $e_j := (0, \, \ldots, \, 0, \, 1, \, \ldots, \, 0) \in
\mathbb{Z}^N$ with the 1 standing at the $j$th position, the $a_j$ 
are bounded complex-valued functions on $\mathbb{Z}^N$, and $\Phi$ is a 
bounded real-valued function on $\mathbb{Z}^N$. The vector $a := 
(a_1, \, \ldots , \, a_N)$ can be viewed of as the discrete analog of 
the magnetic potential, whereas $\Phi$ serves as the discrete analog of 
the electric potential. Clearly, $H$ acts as a self-adjoint operator 
on the Hilbert space $l^2 (\mathbb{Z}^N)$.
\subsection{Slowly oscillating potentials}
We suppose that the vector potential $a = (a_1, \, \ldots, \, a_N)$
and the scalar potential $\Phi$ are slowly oscillating at
infinity. Moreover, we assume that
\begin{equation}  \label{e4.1}
\lim_{x \to \infty} |a_j (x)| = 1 \quad \mbox{for each} \; j = 1, \,
\ldots, \, N.
\end{equation}
Under these assumptions, all limit operators $H_g$ of $H$ are of the
form
\[
H_g = \sum_{k=1}^N (V_{e_k} - a_k^g I) \, (V_{-e_k} - \bar{a}_k^g) +
\Phi^g I
\]
where the $a_k^g$ are complex constants and the $\Phi^g$ are real 
constants given by the pointwise limits
\[
a_k^g = \lim_{m \to \infty} a_k (x + g(m)) \quad \mbox{and} \quad
\Phi^g = \lim_{m \to \infty} \Phi (x + g(m)).
\]
Moreover, $|a_k^g| = 1$. Hence, the limit operators $H_g$ can be 
written as Schr\"{o}dinger operators of the form
\[
H_g = \sum_{k=1}^N (2 - a_k^g V_{-e_k} - \bar{a}_k^g V_{e_k}) + \Phi^g I.
\]
Write $a_k^g$ as $e^{i \varphi_k^g}$ with $\varphi_k^g \in [0, \, 2\pi)$
and set $\varphi^g = (\varphi_1^g, \, \ldots, \, \varphi_N^g)$.
Consider the unitary operator $U : l^2 (\mathbb{Z}^N) \to
l^2 (\mathbb{Z}^N)$ defined by
\[
(Uv)(x) := e^{-i \langle \varphi^g, \, x \rangle } v(x).
\]
Then
\[
\tilde{H}_g := U H_g U^* =\sum_{k=1}^N (2I - V_{-e_k} - V_{e_k}) + \Phi^g I.
\]
The operator $\tilde{H}_g$ is unitarily equivalent to the operator
of multiplication by the function
\[
[0, \, 2\pi)^N \to \mathbb{C}, \quad (\psi_1, \, \ldots, \, \psi_N)
\mapsto 4 \sum_{k=1}^N \sin^2 \frac{\psi_k}{2} + \Phi^g,
\]
thought of as acting on $L^2 (\mathbb{T}^N)$. Hence,
\begin{equation} \label{4.2}
sp \, (\tilde{H}_g) = \left\{ 4 \sum_{k=1}^N \sin^2 \frac{\psi_k}{2} 
+ \Phi^g : \psi_k \in [0, \, 2\pi) \right\} = [\Phi^g, \, \Phi^g + 4N].
\end{equation}
Since the set of all partial limits of a slowly oscillating function
is connected (see Thorem 2.4.7 in \cite{RRSB} or \cite{RRActa}, for 
instance \cite{RRActa}), formula (\ref{2.4'}) implies
\begin{equation} \label{4.3}
sp_{ess} \, (H) = \bigcup \, [\Phi^g, \, \Phi^g + 4N] =
[m(\Phi), \, M(\Phi) + 4N]
\end{equation}
where the union is taken over all sequences $g$ for which the limit
operator $H_g$ exists, and where
\[
m(\Phi) := \liminf_{x \to \infty} \Phi(x) \quad \mbox{and} \quad
M(\Phi) := \limsup_{x \to \infty} \Phi(x).
\]
Hence, {\em the essential spectrum of the Schr\"{o}dinger operator
$(\ref{4.1})$ does not depend on slowly oscillating magnetic
potentials.}
\subsection{Periodic and semi-periodic potentials}
Next we consider the Schr\"{o}dinger operator $H$ in (\ref{4.1})
with $r$-periodic coefficients, that is, the $a_k$ as well as the 
potential $\Phi$ are $r$-periodic functions. The unitary operator 
$T_r : l^2 (\mathbb{Z}^N) \to l^2 (\mathbb{Z}^N, \, \mathbb{C}^d)$ 
defined by (\ref{f2.5}) induces a unitary equivalence between the 
operator $H$ and the matrix operator 
\[
\mathcal{H} := \sum_{k=1}^N (\Lambda_{e_k} - \mu (a_k)) \,
(\Lambda_{-e_k} - \overline{\mu (a_k)}) + \mu (\Phi),
\]
acting on the space $l^2 (\mathbb{Z}^N, \, \mathbb{C}^d)$. Again,
the $\Lambda_{e_k}$ are defined by (\ref{f2.6}), and the $\mu
(a_k)$ and $\mu (\Phi)$ are given by (\ref {2.5'}). The operator
$\mathcal{H}$ on its hand is unitarily equivalent to the operator
of multiplication by the Hermitian matrix function
\[
\sigma_{\mathcal{H}} (t) = \sum_{k=1}^N (\hat{\Lambda}_{e_k} (t_k) -
\mu (a_k)) \, (\hat{ \Lambda}_{-e_k} (t_k) - \overline{\mu (a_k)}) +
\mu (\Phi), \quad t \in \mathbb{T}^N,
\]
acting on the space $L^2 (\mathbb{T}^N, \, \mathbb{C}^d)$. Let
$\lambda_1^H (t) \le \ldots \le \lambda_d^H (t)$ denote the
eigenvalues of the matrix $\sigma_{\mathcal{H}} (t)$. The
functions $t \mapsto \lambda_k^{\mathcal{H}} (t)$ are continuous on 
$\mathbb{T}^N$ (see \cite{Kato}). Since continuous functions map compact
connected sets into compact connected sets again, the sets
\begin{equation} \label{e3.2.1}
\Gamma_k^H := \{ \lambda_k^{\mathcal{H}} (t) : t \in \mathbb{T}^N\}
\end{equation}
are closed subintervals of $\mathbb{R}$. Hence, {\em the spectrum of 
the periodic Schr\"{o}dinger operator $(\ref{4.1})$ coincides with 
its essential spectrum, and it can be represented as the union of
finitely many real intervals.}

Let now $H$ be a Schr\"{o}dinger operator of the form (\ref{4.1})
with semi-periodic coefficients. Thus, the coefficients $a_k$ and 
the potential $\Phi$ belong to $SP_r (\mathbb{Z}^N)$. Then all
limit operators $H^g$ of $H$ are operators with $r$-periodic
coefficients. Hence, the essential spectrum of $H$ is equal to
\[
sp_{ess} \, (H) := \bigcup_{H^g \in op \, (H)} \bigcup_{k=1}^d
\Gamma_k^{H^g}
\]
with the sets $\Gamma_k^{H^g}$ defined by (\ref{e3.2.1}). 

We conclude this section by an example which shows that a 
Schr\"{o}dinger operator with periodic coefficients can have gaps
in its spectrum, whereas a slowly oscillating perturbation of that
operator can close the gaps. More precisely, we consider the 
one-dimensional Schr\"{o}dinger operator
\[
H := V_{-1} + V_1 + (\Phi + \Psi) I
\]
where $\Phi$ is a $2$-periodic real-valued function on
$\mathbb{Z}$ and $\Psi$ belongs to $SO (\mathbb{Z})$. All limit
operators of $H$ are of the form
\[
H^g = V_{-1} + V_1 + (\Phi^g + \Psi^g) I,
\]
where
\[
\Phi^g (x) = \lim_{k \to \infty} \Phi (x + g(k))
\]
is a $2$-periodic real-valued function and
\[
\Psi^g = \lim_{k \to \infty} \Psi (x + g(k))
\]
is a real constant. It follows from $(\ref{f2.4})$ that
\[
sp \, (V_{-1} + V_1 + \Phi^g I) = sp \, (V_{-1} + V_1 + \Phi I).
\]
Hence,
\[
sp \, (H^g) = sp \, (V_{-1} + V_1 + \Phi I) + \Psi^g.
\]
Let $\Psi = 0$ for a moment. The operator $V_{-1} + V_1 + \Phi I$
is unitarily equivalent to the operator of multiplication by the
matrix function
\[
\mathcal{A} (t) := \left(
\begin{array}{cc}
\Phi (0) & 1+t \\
1+t^{-1} & \Phi (1)
\end{array}
\right), \quad t \in \mathbb{T}.
\]
The eigenvalues of $\mathcal{A} (t)$ are the solutions of the
equation
\begin{equation} \label{4.4}
\lambda^2 - (\Phi (0) + \Phi (1)) \lambda + \Phi (0) \Phi (1) -
\gamma (t) = 0
\end{equation}
where $\gamma (t) := 2 + t + t^{-1}$. Let $|\Phi (0) - \Phi (1)|
> 2$. Then $(\ref {4.4})$ has two different solutions
$\lambda_1 (t) < \lambda_2 (t)$ for every $t \in \mathbb{T}$, and
$\lambda_1, \, \lambda_2$ are smooth real-valued functions on
$\mathbb{T}$. Set $m_j := \min_{t \in \mathbb{T}} \lambda_j (t)$
and $M_j := \max_{t \in \mathbb{T}} \lambda_j (t)$ for $j =1, \,
2$. Then
\[
sp \, (V_{-1} + V_1 + \Phi I) = [m_1, \, M_1] \cup [m_2, \, M_2].
\]
Thus, in case $M_1 < m_2$, the operator $V_{-1} + V_1 + \Phi I$ has
a gap $(M_1, \, m_2)$ in its spectrum.

Now consider the slowly oscillating perturbation $H := V_{-1} + V_1 + 
(\Phi + \Psi) I$ of the operator $V_{-1} + V_1 + \Phi I$. The essential 
spectrum of the perturbed operator $H$ is union of the spectra of its 
limit operators $H^g$. It is evident from what has been said above that
\[
sp \, (H^g) = [m_1 + \Psi^g, \, M_1 + \Psi^g] \cup [m_2 + \Psi^g, \,
M_2 + \Psi^g].
\]
Since the set of all partial limits of a slowly oscillating function
is connected, we conclude that
\begin{eqnarray*}
sp_{ess} \, (H) & = & \bigcup_{H^g \in op \, (H)} [m_1 + \Psi^g, \, 
M_1 + \Psi^g] \cup [m_2 + \Psi^g, \, M_2 + \Psi^g] \\
& = & [m_1 + m(\Psi), \, M_1 + M(\Psi)] \cup [m_2 + m(\Psi), \,
M_2 + M(\Psi)],
\end{eqnarray*}
where $m(\Psi) := \liminf_{x \to \infty} \Psi (x)$ and $M(\Psi) :=
\limsup_{x \to \infty} \Psi (x)$. Thus, if 
\[
M(\Psi) - m(\Psi) \ge m_2 - M_1,
\]
then the perturbed operator $H$ has no gaps in its essential
spectrum, and
\[
sp_{ess} \, (H) = [m_1 + m(\Psi), \, M_2 + M(\Psi)]
\]
in this case.
\subsection{Discrete quantum waveguides}
Here we are going to examine the essential spectrum of the
Schr\"{o}dinger operator
\[
H = \sum_{j=1}^N (2I - V_{e_j} - V_{-e_j}) + \Phi I.
\]
under the assumption that the electric potential $\Phi$ has the
form
\[
\Phi (x) := \chi_+ (x_N) \Phi_+ (x) + \chi_0 (x_N) \Phi_0 (x) +
\chi_- (x_N) \Phi_- (x), \quad x \in \mathbb{Z}^N
\]
where $\chi_+, \, \chi_-$ and $\chi_0$ are equal to 1 on the
intervals $(h_2, \infty), \, (-\infty, \, h_1)$ and $[h_1, \,
h_2]$ and equal to 0 outside these intervals, respectively, and
where $\Phi_\pm$ and $\Phi_0$ are function in $SO (\mathbb{Z}^N)$.
Here, $h_1 < h_2$ are previously fixed real numbers.

To describe the essential spectrum of $H$ we consider the limit
operators of $H$ defined by sequences $g = (g^\prime, \, g_N) :
\mathbb{N} \to \mathbb{Z}^{N-1} \times \mathbb{Z}$ tending to
infinity. We have to distinguish between two cases.

In the first case, we assume that the sequence $g = (g^\prime, \,
g_N)$ is such that $g_N \to \pm \infty$. In this case, the limit
operators of $H$ are of the form
\[
H_g^\pm = \sum_{j=1}^N (2 - V_{e_j} - V_{-e_j}) + \Phi_\pm^g I
\]
where the $\Phi_\pm^g := \lim_{j \to \infty} \Phi_\pm (g(j))$ are
real numbers. Consequently,
\[
sp \, (H_g^\pm) = [\Phi_\pm^g, \, \Phi_\pm^g + 4N],
\]
as mentioned in the previous section.

In the second case, we assume that $g^\prime \to \infty$ whereas
$g_N$ is a constant sequence. Then the limit operators of $H$ are
\begin{equation} \label{4.5}
H_g = \sum_{j=1}^N (2 - V_{e_j} - V_{-e_j}) + \Phi^g I
\end{equation}
where the function $\Phi^g$ depends on $x_N$ only. Moreover, this
function is piecewise constant since
\[
\Phi^g (x_N) = \chi_+ (x_N) \Phi_+^g + \chi_0 (x_N) \Phi_0^g +
\chi_- (x_N) \Phi_-^g
\]
with real numbers $\Phi_\pm^g := \lim_{j \to \infty} \Phi_\pm
(g(j))$ and $\Phi_0^g := \lim_{j \to \infty} \Phi_0 (g(j))$.

The operator (\ref{4.5}) is unitarily equivalent to the operator
of multiplication by the operator-valued function $\hat{H}_g :
\mathbb{T}^{N-1} \to L (l^2 (\mathbb{Z}))$ defined by
\[
\hat{H}_g (t^\prime) := \sum_{j=1}^{N-1} (2 - t_j - t_j^{-1}) I +
(2 - V_{e_N} - V_{-e_N}) + \Phi^g I
\]
where $t^\prime = (t_1, \, \ldots, \, t_{N-1})$. It is well-known
(see \cite{RabinD,Teschl}, for instance) that spectrum of the
one-dimensional Jacobi operator
\[
\mathcal{L}_N^g := (2 - V_{e_N} - V_{-e_N}) + \Phi^g I
\]
is the union of its essential spectrum $\Sigma_g := [\Phi_+^g, \,
4 + \Phi_+^g] \cup [\Phi_-^g, \, 4 + \Phi_-^g]$ with a finite set
$\{\lambda_1^g, \, \ldots, \, \lambda_{m(g)}^g \} $ of points in
the discrete spectrum which are located outside $\Sigma_g$. Hence,
\[
sp \, (\mathcal{L}_N^g) = [\Phi_+^g, \, 4N + \Phi_+^g] \, \bigcup \,
[\Phi_-^g, \, 4N + \Phi_-^g] \bigcup_{j=1}^{m(g)} [\lambda_j^g, \,
4 (N-1)].
\]
Set $\mathbb{Z}_+^N := \{ z \in \mathbb{Z}^N : z_N > 0 \}$ and
$\mathbb{Z}_-^N := \mathbb{Z}^N \setminus \mathbb{Z}_+^N$, and
abbreviate
\[
m(\Phi_\pm) := \liminf_{\mathbb{Z}_\pm^N \ni x \to \infty} \Phi_\pm,
\quad
M(\Phi_\pm) := \limsup_{\mathbb{Z}_\pm^N \ni x \to \infty} \Phi_\pm.
\]
Then equality (\ref{2.4'}) implies that $sp_{ess} \, (H)$ is equal to 
\[
[m(\Phi_-), \, M(\Phi_-) + 4N] \, \bigcup \,
[m(\Phi_+), \, M(\Phi_+) + 4N] \bigcup_{g \in \Lambda} 
\bigcup_{j=1}^{m(g)} [\lambda_j^g, \, 4 (N-1)]
\]
where $\Lambda$ refers to the set of all sequences $g = (g^\prime,
\, g_N)$ for which $g^\prime \to \infty$ and $g_N$ is constant.
\subsection{Potentials with an infinite set of discontinuities}
Set $\Delta := 2I - V_{-1} - V_1 \in L(l^2(\mathbb{Z}))$. We consider 
the Schr\"{o}dinger operator
\[
H = \Delta + \Phi I : l^2 (\mathbb{Z}) \to l^2 (\mathbb{Z})
\]
where the potential $\Phi$ takes two real values $a, \, b$ only,
but where $\Phi$ is allowed to have an infinite set of jumps. More
precisely, let $\Phi (x) = a$ if $x \in \Lambda$ and let $\Phi (x)
= b$ if $x \in \mathbb{Z} \setminus \Lambda$ where
\[
\Lambda := \cup_{k=0}^\infty \, \{x \in \mathbb{Z} : \gamma_k^- \le
|x| \le \gamma_k^+ \}
\]
and where the $\gamma_k^-$ and $\gamma_k^+$ are integers satisfying
\begin{equation} \label{4.6'}
\lim_{k \to \infty} \gamma_k^- = \lim_{k \to \infty} \gamma_k^+
= \lim_{k \to \infty} (\gamma_k^+ - \gamma_k^-) = \lim_{k \to \infty}
(\gamma_{k+1}^- - \gamma_k^+) = +\infty.
\end{equation}
For example, this condition holds if $\gamma_k^- := k^2$ and
$\gamma_k^+ := k^2 + k$.
\begin{theorem} \label{t3.1}
Under these assumptions, $sp_{ess} \, (\Delta + \Phi I) = [a, \, a+4]
\cup [b, \, b+4]$.
\end{theorem}
\begin{proof}
First we determine all limit operators of the operator $\Phi I$
with respect to sequences $g$ tending to $+\infty$. Again, we have
to distinguish between several possibilities. \\[2mm]
{\em Case a}) Let the sequence $g \to +\infty$ satisfy
\[
\lim_{k \to \infty} (\gamma_k^+ - g(k)) =
\lim_{k \to \infty} (g(k) - \gamma_k^-) = +\infty.
\]
Such sequences exist because of $\gamma_k^+ - \gamma_k^- \to +\infty$.

In this case, for every $x \in \mathbb{Z}$, there exists a $k(x)$
with $x + g(k) \in (\gamma_k^-, \, \gamma_k^+)$ for each $k \ge
k(x)$. Thus, $\lim_{k \to \infty} \Phi (x + g(k)) = a$ for every
$x \in \mathbb{Z}$, whence
\[
\mbox{s-lim}_{k \to \infty} V_{-g(k)} \Phi V_{g(k)} = aI.
\]
{\em Case b}) Let the sequence $g \to +\infty$ be such that
\[
\lim_{k \to \infty} (\gamma_{k+1}^- - g(k)) =
\lim_{k \to \infty} (g(k) - \gamma_k^+) = +\infty.
\]
Such sequences exist because of $\gamma_{k+1}^- - \gamma_k^+ \to
+\infty$. 

Under this assumption, for every $x \in \mathbb{Z}$, one finds a
$k(x)$ such that $x + g(k) \in (\gamma_k^+, \, \gamma_{k+1}^-)$
for each $k \ge k(x)$. Thus, $\lim_{k \to \infty} \Phi (x + g(k))
= b$ for every $x \in \mathbb{Z}$, whence
\[
\mbox{s-lim}_{k \to \infty} V_{-g(k)} \Phi V_{g(k)} = bI.
\]
{\em Case c}) Let now $g(k) = \gamma_k^- - h(k)$ with a bounded
sequence $h$. Passing to a suitable subsequence, one can assume
that $g(k) = \gamma_k^- - h$ where $h$ is constant. If $x - h <
0$, then $\lim_{k \to \infty} \Phi (x + g(k)) = b$, whereas this
limit is equal to $a$ in case $x - h \ge 0$. We denote the
characteristic function of the set $\mathbb{Z}_+ := \{ 0 \} \cup
\mathbb{N}$ by $\chi_+$ and set $\chi_- := 1 - \chi_+$. Then
\[
\lim_{k \to \infty} \Phi (x + g(k)) = a \chi_+ (x-h) + b \chi_-
(x-h), \quad \mbox{for} \; x \in \mathbb{Z},
\]
which implies that
\[
\mbox{s-lim}_{k \to \infty} V_{-g(k)} \Phi V_{g(k)} =
(a \chi_+^h + b \chi_-^h) I
\]
with $\chi_\pm^h (x) := \chi_\pm (x-h)$. \\[2mm]
{\em Case d}) Let finally $g(k) = \gamma_k^+ - h(k)$ where $h$ is
a bounded sequence. Passing to a suitable subsequence one can
assume again that $g(k) = \gamma_k^- - h$ where $h$ is constant.
Now one has $\lim_{k \to \infty} \Phi (x + g(k)) = a$ for $x - h <
0$, and this limit is equal to $b$ for $x - h \ge 0$. Consequently,
\[
\lim_{k \to \infty} \Phi (x + g(k)) = a \chi_-^h (x) + b \chi_+^h
(x), \quad \mbox{for} \; x \in \mathbb{Z}
\]
with $\chi_\pm^h$ as above. This shows that
\[
\mbox{s-lim}_{k \to \infty} V_{-g(k)} \Phi V_{g(k)} = (a \chi_-^h +
b \chi_+^h) I.
\]
Evidently, every sequence $g$ tending to $+ \infty$ is subject to
one of these four cases. Thus, the set of the limit operators of
$\Phi I$ defined by sequences tending to $+\infty$ is exhausted by
the operators $aI, \, bI, \, (a \chi_+^h + b \chi_-^h) I$ and $(a
\chi_-^h + b \chi_+^h) I$ where $h$ runs through $\mathbb{Z}$.
Similarly one checks that the set of all limit operators of $\Phi
I$ defined by sequences $g$ tending to $-\infty$ consists exactly
of the same operators $aI, \, bI, \, (a \chi_+^h + b \chi_-^h) I$
and $(a \chi_-^h + b \chi_+^h) I$. Notice further that the
operators $(a \chi_+^h + b \chi_-^h) I$ and $(a \chi_-^h + b
\chi_+^h) I$ are unitarily equivalent to the operators $(a
\chi_+^0 + b \chi_-^0) I = (a \chi_+ + b \chi_-) I$ and $(a
\chi_-^0 + b \chi_+^0) I = (a \chi_- + b \chi_+) I$, respectively.

It is evident that the continuous spectra of the operators $\Delta
+ aI$ and $\Delta + bI$ with constants $a, \, b$ are the intervals
$[a, \, a+4]$ and $[b, \, b+4]$, respectively, whereas both the
spectrum of the operator $\Delta + (a \chi_+ + b \chi_-)I$ and
that of $\Delta + (a \chi_- + b \chi_+) I$ are equal to $[a, \,
a+4] \cup [b, \, b+4]$ (see \cite{RabinD}, for instance). Hence,
the assertion follows from Theorem \ref{t2.1}.
\end{proof} \\[3mm]
The method of limit operators can be also applied to more
involved potentials $\Phi$ of the form
\begin{equation} \label{4.7'}
\Phi = a \chi_\Lambda + b \chi_{\mathbb{Z} \setminus \Lambda}
\end{equation}
where $\chi_\Lambda$ is the characteristic function of a subset
$\Lambda$ of $\mathbb{Z}$ as before, and where $a, \, b \in SO
(\mathbb{Z})$.
\begin{theorem} \label{t3.2}
Let $\Phi$ be a potential of the form $(\ref{4.7'})$. Then
\begin{equation} \label{f4.8}
sp_{ess} \, (\Delta + \Phi I) = [m(a), \, M(a) + 4] \cup [m(b), \,
M(b) + 4]
\end{equation}
where $m(f) := \liminf_{x \to \infty} f(x)$ and $M(f) :=
\limsup_{x \to \infty} f(x)$ for each slowly oscillating function
$f$.
\end{theorem}
\begin{proof}
We proceed as in the proof of the previous theorem to find that
all limit operators of $\Phi I$ are unitarily equivalent to one
of the operators
\[
a^g I, \; b^g I, \; (a^g \chi_- + b^g \chi_+) I, \; (a^g \chi_+ 
+ b^g \chi_-) I
\]
with constants
\begin{equation} \label{e3.4.2}
a^g := \lim_{k \to \infty} a(g(k)) \quad \mbox{and} \quad
b^g := \lim_{k \to \infty} b(g(k)).
\end{equation}
Thus, according to Theorems \ref{t3.1} and \ref{t2.1},
\[
sp_{ess} \, (H) = \bigcup ([a^g, \, a^g + 4] \cup [b^g, \, b^g + 4])
\]
where the union is taken over all sequences $g \to \infty$ for
which the partial limits (\ref{e3.4.2}) of the functions $a$ and 
$b$ exist. Employing the connectedness of the set of all partial
limits of a slowly oscillating function, we arrive at identity
(\ref{f4.8}).
\end{proof}
\subsection{The three-particle problem}
We consider the three-particle Schr\"{o}dinger operator on
$l^2 (\mathbb{Z}^3 \times \mathbb{Z}^3)$
\begin{eqnarray} \label{4.6}
H & := & \frac{1}{2m_1} (\Delta \otimes I) + \frac{1}{2m_2} (I
\otimes \Delta) \\
&& \quad + \; (W_1 I) \otimes I + I \otimes (W_2 I) +
W_{12}^{dif} (I \otimes I)  \nonumber
\end{eqnarray}
which describes the motion on the lattice $\mathbb{Z}^3$ of two
particles with coordinates $x^1, \, x^2 \in \mathbb{Z}^3$ with
masses $m_1, \, m_2$ around a heavy nuclei located at the point
$0$. In (\ref{4.6}) we write $I$ for the identity operator on $l^2
(\mathbb{Z}^3)$ and $\Delta$ for the discrete Laplacian
\[
\Delta := \sum_{j=1}^3 (2 I - V_{-e_j} - V_{e_j})
\]
on $l^2 (\mathbb{Z}^3)$ where $\{ e_1, \, e_2, \, e_3 \}$ stands
for the standard basis of $\mathbb{Z}^3$. Further, $W_1, \, W_2,
\, W_{12}$ are real-valued functions on $\mathbb{Z}^3$ with
\[
\lim_{z \to \infty} W_1 (z) = \lim_{z \to \infty} W_2 (z) =
\lim_{z \to \infty} W_{12} (z) = 0,
\]
and $W_{12}^{dif}$ is the function on $\mathbb{Z}^3 \times
\mathbb{Z}^3$ defined by
\[
W_{12}^{dif} (x_1, \, x_2) := W_{12} (x_1 - x_2).
\]
Further we abbreviate $m := 6/m_1 + 6/m_2$.

In order to describe the essential spectrum of the operator $H$ by
means of (\ref{2.4'}), we have to determine the limit operators of 
$H$. Let $g = (g^1, \, g^2) : \mathbb{N} \to \mathbb{Z}^3 \times 
\mathbb{Z}^3$ be a sequence tending to infinity. The examination of
the limit operators leads to the consideration of the following 
situations. \\[2mm]
{\em Case 1}) The sequence $g^1$ tends to infinity, whereas $g^2$
is constant. Then the limit operator $H_g$ of $H$ is unitarily
equivalent to the operator
\begin{equation} \label{e4.7}
H_2 := \frac{1}{2m_1} (\Delta \otimes I) + \frac{1}{2m_2} 
(I \otimes \Delta) + I \otimes (W_2 I).
\end{equation}
{\em Case 2}) If $g^2 \to \infty$ and $g^1$ is constant, then the
limit operator $H_g$ of $H$ is unitarily equivalent to the
operator
\begin{equation} \label{e4.8}
H_1 := \frac{1}{2m_1} (\Delta \otimes I) + \frac{1}{2m_2}
(I \otimes \Delta) + (W_1 I) \otimes I.
\end{equation}
{\em Case 3a}) Let both $g^1 \to \infty$ and $g^2 \to \infty$,
and assume that also $g^1 - g^2 \to \infty$. In this case, the
associated limit operator of $H$ is the free discrete Hamiltonian
$\frac{1}{2m_1} (\Delta \otimes I) + \frac{1}{2m_2} (I \otimes \Delta)$. 
The spectrum of this operator is the intervall $[0, \, m]$. \\[2mm]
{\em Case 3b}) Let again $g^1 \to \infty$ and $g^2 \to \infty$,
but now let $g^1 - g^2$ be a bounded sequence. Then $g^1 - g^2$
hac a constant subsequence; so we can assume without loss that the
sequence $g^1 - g^2$ itself is constant. In this case, the limit
operator with respect to $g$ is unitarily equivalent to the
operator of interaction of the particles $x^1, \, x^2$ which is
given by
\begin{equation} \label{4.10}
H_{12} := \frac{1}{2m_1} (\Delta \otimes I) + \frac{1}{2m_2}
(I \otimes \Delta) + W_{12}^{dif} (I \otimes I).
\end{equation}
It follows from Proposition \ref{pr2.1} that the discrete spectra
of the operators $H_1, \, H_2$ and $H_{12}$ are empty. Thus, since
the spectrum of each operator $H_1, \, H_2$ and $H_{12}$ contains 
the interval $[0, \, m]$ with $m$ as above, and due to (\ref{2.4'}),
\begin{equation} \label{4.11}
sp_{ess} \, (H) = sp \, (H_1) \cup sp \, (H_2) \cup sp \, (H_{12}).
\end{equation}
To determine the spectrum of $H_2$, we apply the Fourier transform
with respect to the first variable. Then $H_2$ becomes unitarily
equivalent to the operator of multiplication by the
operator-valued function
\begin{equation} \label{4.7}
\mathbb{T}^3 \to L(l^2(\mathbb{Z}^3)), \quad t \mapsto
\sum_{j=1}^3 \frac{1}{2m_1} (2 - t_j -t_j^{-1}) I + \frac{1}{2m_2}
\Delta + W_2 I.
\end{equation}
The operator $\frac{1}{2m_2} \Delta + W_2 I$ acting on 
$l^2 (\mathbb{Z}^3)$ has the essential spectrum 
\[
\frac{1}{2m_2} [0, \, 12] = [0, \, 6/m_2]
\]
and a real discrete spectrum
$\{ \lambda_k^{(2)} \}_{k=1}^\infty$ which is located outside the
interval $[0, \, 6/m_2]$ and which has $0$ and $6/m_2$ as only
possible accumulation points. Consequently,
\begin{equation} \label{4.8}
sp \, (H_2) = [0, \, m] \bigcup_{k=1}^\infty 
[\lambda_k^{(2)}, \, \lambda_k^{(2)} + 6/m_2].
\end{equation}
In the same way one obtains
\begin{equation} \label{4.8'}
sp \, (H_1) = [0, \, m] \bigcup_{k=1}^\infty 
[\lambda_k^{(1)}, \, \lambda_k^{(1)} + 6/m_1]
\end{equation}
where $\{ \lambda_k^{(1)} \}_{k=1}^\infty$ is the sequence of the
points of the discrete spectrum of $\frac{1}{2m_1} \Delta + W_1 I$ 
on $l^2 (\mathbb{Z}^3)$ which are located outside $[0, \, 6/m_1]$ 
and which can accumulate only at $0$ and $12$.

Note that, unlike the continuous case of operators on
$\mathbb{R}^6$, there is no transformation of the discrete
operator $H_{12}$ to an operator of the form $\Delta \otimes I + I
\otimes \Delta + (W_{12} I) \otimes I$ with a multiplication
operator $W_{12} I$ on $l^2(\mathbb{Z}^3)$. This fact makes the
spectral theory of discrete operators of interaction much more
difficult than the corresponding theory for continuous operators
of interaction on $\mathbb{R}^3$ (see, for instance,
\cite{AlbaverioLakaev,AlbLakMum,LakaevMuminov}).

We can only give a simple estimate for the location of the
spectrum of $H_{12}$ which results from the following well-known
estimate for the spectrum of a self-adjoint operator acting on a
Hilbert space (\cite{Lax}, p. 357).
\begin{proposition} \label{Pr3.1}
Let $A$ be a self-adjoint operator on a Hilbert space. Then $sp \, (A) 
\subseteq [a, \, b]$ where
\[
a := \inf_{\|h\| = 1} \langle Ah, \, h \rangle, \quad b :=
\sup_{\|h\| = 1} \langle Ah, \, h \rangle.
\]
The points $a, \, b$ belong to the spectrum of $A$.
\end{proposition}
Proposition \ref{Pr3.1} implies the following estimates for the 
spectra of $H_1, \, H_2$ and $H_{12}$:
\[
[0, \, m] \subseteq sp \, (H_j) \subseteq 
\left[ \inf_{x \in \mathbb{Z}^3} W_j (x), \, 
\sup_{x \in \mathbb{Z}^3} W_j (x) + m \right]
\]
for $j=1, \, 2$, and
\[
[0, \, m] \subseteq sp \, (H_{12}) \subseteq 
\left[ \inf_{x \in \mathbb{Z}^3} W_{12} (x), \, 
\sup_{x \in \mathbb{Z}^3} W_{12} (x) + m \right].
\]
In combination with (\ref{4.11}), these inclusions yield lower and
upper bounds for the essential spectrum of $H$.
\begin{theorem}
Let $H$ be the Schr\"{o}dinger operator $(\ref{4.6})$ on 
$l^2 (\mathbb{Z}^3 \times \mathbb{Z}^3)$. Then
\[
\inf sp_{ess} \, (H) = \min \left( \inf_{x \in \mathbb{Z}^3} W_1(x), 
\, \inf_{x \in \mathbb{Z}^3} W_2(x), \, \inf_{x \in \mathbb{Z}^3}
W_{12}(x) \right),
\]
\[
\sup sp_{ess} H = \max \left( \sup_{x \in \mathbb{Z}^3} W_{1}(x) +
m, \, \sup_{x \in \mathbb{Z}^3} W_2(x) + m, \, \sup_{x \in
\mathbb{Z}^3} W_{12}(x) + m \right)
\]
with $m := 6/m_1 + 6/m_2$.
\end{theorem}
{\small Authors' addresses: \\[3mm]
Vladimir S. Rabinovich, Instituto Polit\'{e}cnico Nacional, \\
ESIME-Zacatenco, Av. IPN, edif. 1, M\'{e}xico D.F., 07738, M\'{E}XICO. \\
e-mail: vladimir$_-$rabinovich@hotmail.com \\[1mm]
Steffen Roch, Technische Universit\"{a}t Darmstadt, \\
Schlossgartenstrasse 7, 64289 Darmstadt, Germany.\\
e-mail: roch@mathematik.tu-darmstadt.de}
\end{document}